
%
%
%
%
%
%
%
\mag=\magstep1
\hsize=5.5in
\vsize=6.9in
\baselineskip=14pt plus 2pt
%
%
%
%
%
%
\catcode`@=11 
%
%
%
%
%
\def\titleline#1{\centerline{\bf #1}\par\vskip 1cm}
\def\abstract{\par\penalty-100\medskip
   \spacecheck\sectionminspace
   \line{\tenrm \hfil Abstract \hfil}
   \nobreak\vskip\headskip }

\def\chcont#1#2{\line{\hbox to 15pt{\hss{\bf #1. }} #2 \hss}}
\def\seccont#1#2{\line{\hbox to 30pt{\hss#1. } #2 \hss}}

\def\leaderfill{\leaders\hbox to 1em{\hss.\hss}\hfill}
%
%
%
%
\newcount\chapternumber	     \chapternumber=0
\newcount\sectionnumber	     \sectionnumber=0
\newskip\chapterskip	     \chapterskip=\bigskipamount
\newskip\sectionskip	     \sectionskip=\medskipamount
\newskip\headskip	     \headskip=8pt plus 3pt minus 3pt
\newdimen\chapterminspace    \chapterminspace=15pc
\newdimen\sectionminspace    \sectionminspace=10pc
\def\spacecheck#1{\dimen@=\pagegoal\advance\dimen@ by -\pagetotal
   \ifdim\dimen@<#1 \ifdim\dimen@>0pt \vfil\break \fi\fi}
\def\titlestyle#1{\par\begingroup
   \interlinepenalty=9999
   \leftskip=0.03\hsize plus 0.20\hsize minus 0.03\hsize
   \rightskip=\leftskip    \parfillskip=0pt
   \hyphenpenalty=9000     \exhyphenpenalty=9000
   \tolerance=9999         \pretolerance=9000
   \spaceskip=0.333em      \xspaceskip=0.5em
   \tenbf  \noindent  #1\par\endgroup }
\def\chapter#1{\par \penalty-300 \vskip\chapterskip
   \spacecheck\chapterminspace
   \chapterreset \titlestyle{\chapterlabel \ #1}
   \nobreak\vskip\headskip \penalty 30000}
\def\chapterreset{\global\advance\chapternumber by 1
   \ifnum\equanumber<0 \else\global\equanumber=0\fi
   \sectionnumber=0
   \resetregist
   \xdef\chapterlabel{\number\chapternumber.}}
\def\resetregist{}
\def\alphabetic#1{\count255='140 \advance\count255 by #1\char\count255}
\def\Alphabetic#1{\count255='100 \advance\count255 by #1\char\count255}
\def\Roman#1{\uppercase\expandafter{\romannumeral #1}}

\def\section#1{\par \ifnum\the\lastpenalty=30000\else
   \penalty-200\vskip\sectionskip \spacecheck\sectionminspace\fi
   \global\advance\sectionnumber by 1  \noindent
   {\bf \enspace \chapterlabel \sectionlabel #1} \par
   \nobreak \vskip\headskip \penalty 30000 }
\def\sectionlabel{\number\sectionnumber \quad }
\def\subsection#1{\par
   \ifnum\the\lastpenalty=30000\else \penalty-100\smallskip \fi
   \noindent\undertext{#1}\enspace \vadjust{\penalty5000}}
\def\undertext#1{\vtop{\hbox{#1}\kern 1pt \hrule}}
\def\unnumberedchapters{\let\makel@bel=\relax \let\chapterlabel=\relax
   \let\sectionlabel=\relax \equanumber=-1 }
%

%
%
%
%
\newcount\referencecount     \referencecount=0
\newdimen\referenceminspace  \referenceminspace=25pc
\newif\ifreferenceopen	     \newwrite\referencewrite
\newdimen\refindent	\refindent=30pt
\def\REF#1#2{\refch@ck
   \global\advance\referencecount by 1 \xdef#1{\the\referencecount}
   \immediate\write\referencewrite{%
   \noexpand\refitem{[#1]}
   #2}}
\def\refch@ck{\chardef\rw@write=\referencewrite \ifreferenceopen \else
   \referenceopentrue
   \immediate\openout\referencewrite=reference.aux \fi}
\def\refitem#1{\par \hangafter=0 \hangindent=\refindent \Textindent{#1}}
\def\Textindent#1{\noindent\llap{#1\enspace}\ignorespaces}
\def\NP{Nucl.\ Phys.\ }
\def\PR{Phys.\ Rev.\ }
\def\PL{Phys.\ Lett.\ }

\def\PTP{Prog.\ Theor.\ Phys.\ }
%
%
%
%
\newcount\equanumber	     \equanumber=0
\def\eqalignnop#1{\relax \ifnum\equanumber<0
   \eqalignno{#1}\global\advance\equanumber by -1
   \else\global\advance\equanumber by 1 \eqalignno{#1} \fi}
\def\eqname#1{\relax \ifnum\equanumber<0
   \xdef#1{{\rm(\number-\equanumber)}}\global\advance\equanumber by -1
   \else \global\advance\equanumber by 1
   \xdef#1{{\rm(\chapterlabel \number\equanumber)}} \fi}
\def\eqnamep#1{\relax \ifnum\equanumber<0
   \xdef#1##1{{\rm(\number-\equanumber##1)}}
   \else \xdef#1##1{{\rm(\chapterlabel \number\equanumber##1)}} \fi}
\def\eqn#1{\eqno\eqname{#1}#1}

%

%
%
%
%
\newcount\lastf@@t	     \lastf@@t=-1
\newcount\footsymbolcount    \footsymbolcount=0
\let\footsymbol=\star
\def\footsymbolgen{\relax
   \NPsymbolgen 
   \global\lastf@@t=\pageno}
\def\NPsymbolgen{\ifnum\footsymbolcount<0 \global\footsymbolcount=0\fi
   {\advance\lastf@@t by 1
   \ifnum\lastf@@t<\pageno \global\footsymbolcount=0
   \else \global\advance\footsymbolcount by 1 \fi }
   \ifcase\footsymbolcount \fd@f\star\or \fd@f\dagger\or \fd@f\ast\or
   \fd@f\ddagger\or \fd@f\natural\or \fd@f\diamond\or \fd@f\bullet\or
   \fd@f\nabla\else \fd@f\dagger\global\footsymbolcount=0 \fi }
\def\fd@f#1{\xdef\footsymbol{#1}}
\def\PRsymbolgen{\ifnum\footsymbolcount>0 \global\footsymbolcount=0\fi
   \global\advance\footsymbolcount by -1
   \xdef\footsymbol{\sharp\number-\footsymbolcount} }
%
%
%
%
\def\pagebreak{\vfil\eject}

\def\units#1{\,{\rm #1}}
\def\lsim{\mathrel{\mathpalette\@versim<}}
\def\gsim{\mathrel{\mathpalette\@versim>}}
\def\@versim#1#2{\lower0.2ex\vbox{\baselineskip\z@skip\lineskip\z@skip
  \lineskiplimit\z@\ialign{$\m@th#1\hfil##\hfil$\crcr#2\crcr\sim\crcr}}}
%
%
%
%
%
\unnumberedchapters
%
%
%
%
\def\gS{\Sigma}
\def\bk{{\bf k}}
\def\Nb{{\bar N}}
\def\cO{{\cal O}}
\def\cL{{\cal L}}
\def\r{\right}
\def\l{\left}
\def\gSp{\gS_{\pi N}}
\def\fp{f_\pi}
\def\mp{m_\pi}
\def\TpN{T_{\pi N}}
\def\Box{\mathop{\lower0.2ex
         \vbox{\hrule\hbox{\vrule
           \vrule width0pt depth0pt height1.2ex
           \vrule height0pt depth0pt width1.2ex
                                   \vrule}\hrule}}}
\def\JOUR#1#2#3#4{#1 {\bf #2}, #4 (#3)}
\def\bra<#1|{\left\langle #1\right\vert}
\def\ket|#1>{\left\vert #1\right\rangle}
\def\VEV<#1>{\left\langle #1\right\rangle}
\def\Ip<#1|#2>{{\la #1 | #2 \ra}}
\def\Ex<#1|#2|#3>{{\langle #1 | #2 | #3 \rangle}}
%
%
%
%
\REF\KAP{D.~B. Kaplan and A.~E. Nelson,
                       \JOUR{\PL}{B175}{1986}{57};
                       \JOUR{}{B179}{1986}{409(E)}.}
\REF\NEL{A.~E. Nelson and D.~B. Kaplan,
                      \JOUR{\PL}{B192}{1987}{193}.}
\REF\BKR{G.~E. Brown, K. Kubodera and M. Rho,
                      \JOUR{\PL}{B192}{1987}{273}.}
\REF\PW{H.~D. Politzer and M.~B. Wise,
                      \JOUR{\PL}{B273}{1991}{156}.}
\REF\BKRT{G.~E. Brown, K. Kubodera, M. Rho and V. Thorsson,
                       \JOUR{\PL}{B291}{1992}{355}.}
\REF\TT{T. Tatsumi, \JOUR{\PTP}{80}{1988}{22}, \hfill\break
        T. Muto and T. Tatsumi, \JOUR{\PL}{B283}{1992}{165}.}
\REF\BLRT{G.~E. Brown, C.-H. Lee, M. Rho and V. Thorsson,
                      Nucl. Phys. A, to be published.}
\REF\LJMR{C.-H. Lee, H. Jung, D.-P. Min and M. Rho,
                      preprint SNUTP-93-81.}
\REF\DEE{J. Delorme, M. Ericson and T.~E.~O. Ericson,
                         \JOUR{\PL}{B291}{1992}{379}.}
\REF\YNK{H. Yabu, S. Nakamura and K. Kubodera,
                         \JOUR{\PL}{B317}{1993}{269}.}
\REF\YNMK{H. Yabu, S. Nakamura, F. Myhrer and K. Kubodera,
                         \JOUR{\PL}{B315}{1993}{17}.}
\REF\LUTa{M. Lutz, A. Steiner and W. Weise,
                         \JOUR{\PL}{B278}{1992}{29}.}
\REF\LUTb{M. Lutz, A. Steiner and W. Weise,
                         Regensburg preprint,1993.}
\REF\ADLa{For review, see S.~L. Adler and R.~F. Dashen,
    {\it Current Algebra and Applications to Particle Physics},
         (Benjamin, 1968).}
\REF\BRO{L.~S. Brown, W.~J. Pardee and R.~D. Peccei,
                      \JOUR{\PR}{D4}{1971}{2801}.}
\REF\GEL{M. Gell-Mann, R.~J. Oakes and B. Renner,
                      \JOUR{\PR}{175}{1968}{2195}.}
\REF\COMa{
One might think that the first term in eq.{\rm (8)}
is singular in the chiral limit $\mp$=0.
However, using the Gell-Mann-Oaks-Renner relation \hfill\break
$\fp^2\mp^2={1 \over 2} (m_u+m_d)
\Ex<0| {\bar u} u+{\bar d} d |0>$ [\GEL]
and eq. {\rm (2)}, 
one can see that, as $\mp \rightarrow 0$,
eq. {\rm (8)} 
tends to the finite limit
$$
     \TpN^{(2)} ={\Ex<N| {\bar u} u +{\bar d} d|N>
              \over \Ex<0| {\bar u} u +{\bar d} d|0>}
                 (k^2+(k')^2)  +\TpN'|_{\mp = 0}.
$$}
\REF\COMb{
The importance of matter effects for the Born term
in the context of the Nambu-Jona-Lasinio model
was discussed by Lutz et al. [\LUTa].}
\REF\LEU{J. Gasser and H. Leutwyler, \JOUR{Ann. Phys.}
       {158}{1984}{142}.}
\REF\COMc{
It is known [\LEU] that
by working with a chiral effective Lagrangian
with gauged external source terms
one can recover the Adler condition in ChPT.
This procedure, however, amounts to using
$\pi$ of eq. {\rm (6)} 
as the pion field operator
instead of the original field
that appears in the effective Lagrangian.
Therefore, this ``reconciliation" between ChPT and PCAC
does not solve the problem we are facing here.}
\REF\ER{D. Ebert and H. Reinhardt,
\JOUR{\NP}{B271}{1986}{188}.}
\REF\RCSI{C.~D. Roberts, R.T. Cahill, M.E. Sevior
          and N. Iannella, \JOUR{\PR}{D49}{1994}{125}.}
\REF\COMd{
Lagrangians containing box terms
have been used in, e.g.,
[\ER, \RCSI].
In ChPT, the box terms in the pure mesonic sector
can be eliminated simply by redefining the meson field.}
\REF\GSS{J. Gasser, M.~E. Sainio and A.\u{S}varc,
        \JOUR{\NP}{B307}{1988}{779}.}
\REF\COMe{
This also allows us to circumvent the mathematical
subtleties
associated with the treatment of operator products.}
\REF\COMf{
For instance, the QED Lagrangian
is bilinear in Fermion fields.}
\REF\WEI{S. Weinberg, \JOUR{\PL}{B251}{1990}{288};
        \JOUR{\NP}{B363}{1991}{3}.}
\REF\COMg{
Usual ChPT tests involving single baryons
place no constraints on these \hfill\break
``non-standard" terms.
For the ordinary chiral counting,
a vertex with a larger number of fermion lines
leads to a higher chiral index $\nu_i$
and hence its contribution is suppressed [\WEI].
The issue here is whether this ``ordinary" counting
can be applied to nuclear matter,
which has an additional dimensional parameter,
$k_F$ or $\rho$.
We shall come back to this point later in the text.}
\REF\BKM{The one-loop corrections
     for the pion-nucleon scattering lengths were
     calculated in: V. Bernard, N. Kaiser and U-G. Meissner,
     \JOUR{\PL}{B309}{1993}{421}.}
\REF\COMh{
In a ChPT calculation of
the nucleon-nucleon interactions by van Kolck ,
the multiple-fermion terms play an essential role:
U. L. van Kolck, thesis (University of Texas, 1993),
unpublished.}
\REF\COMi{
We are grateful to Ubirajara van Kolck for the discussion
on this point.}
\REF\COMj{
In general, from the low energy theorem
for the $N^n(\pi,\pi)N^n$ process,
the scattering amplitude at the soft point is given
by the generalized sigma term, \hfill\break
$\gS_{\pi N^n} ={1 \over 2} (m_u+m_d)
                \Ex<N^n| {\bar u} u +{\bar d} d |N^n>,$
which might be estimated in some models.}
\REF\COMk{
The influence of the $b$ term of the s-wave $\TpN$
was studied by Delorme et al. [\DEE]
along with some other $\cO(\rho^2)$ effects.}
\REF\BB{G.E. Brown and H.A. Bethe, Astrophys. J.,
to be published.}
%
%
%
%
\noindent{USC(NT)--94--1}\par
\vskip 1cm
\centerline{{\bf Meson Condensation
in Dense Matter Revisited}\footnote{*}{
Supported in part by the NSF under Grant
No. PHYS-9310124}}
\vskip 1.5cm
\centerline{Hiroyuki Yabu, F. Myhrer, and K. Kubodera}
\vskip0.8cm
\centerline{Department of Physics and Astronomy}
\centerline{University of South Carolina}
\centerline{Columbia, South Carolina 29208, USA}
\vskip 2cm
\titleline{Abstract}

The results for meson condensation in the literature
vary markedly depending on
whether one uses chiral perturbation theory
or the current-algebra-plus-PCAC approach.
To elucidate the origin of this discrepancy,
we re-examine the role of the sigma-term
in meson condensation.
We find that the resolution of the existing discrepancy
requires a knowledge of terms in the Lagrangian
that are higher order in density than hitherto considered.
\pagebreak
%
%
%
%
%
Kaon condensation in dense nuclear matter was proposed
some time ago by Kaplan and Nelson
[\KAP],
who used a particular effective Lagrangian
derived from chiral perturbation theory (ChPT)
and employed the tree approximation.
In this Lagrangian
the attractive force that drives condensation
is provided primarily by the $K$-$N$ sigma term,
which is expected to be much larger
than the $\pi$-$N$ sigma term.
For $\gS_{KN} = (400 \sim 600) \units{MeV}$,
the critical density for kaon condensation
was predicted to be
$\rho_c = \,2 \sim 3 \, \rho_0$
($\rho_0$ = normal nuclear density) [\KAP].
This remarkable result gave strong impetus to
further detailed studies
of kaon condensation and its possible influences
on neutron stars [\NEL-\LJMR].
Along the line of ChPT,
a systematic examination of higher order terms
in chiral expansion has been pursued
using the heavy-fermion formalism
[\BKRT,\BLRT,\LJMR].
Meanwhile,
several authors
[\DEE-\LUTb]
have recently questioned
the validity of kaon condensation
driven by the $K$-$N$ sigma term.
In particular, Yabu et al. [\YNMK]
demonstrated explicitly that
the use of $K$-$N$ scattering amplitudes
that respect the current algebra theorems and PCAC
does not lead to kaon condensation.
An important question is
why the {\it existing} calculations on kaon condensation
give markedly different results
depending on whether one uses ChPT
or the current algebra approach.
In this note we analyze
the nature of the problem involved
and discuss what kind of additional information
is required to settle the issue.
Since the sigma term is
the central issue here, we first concentrate our attention
on the role of the sigma term.
Furthermore, for the illustrative purpose,
we consider s-wave pion condensation
rather than kaon condensation itself.
We will argue that terms of $\cO (\rho^2)$
in the Lagangian are of importance
to resolve the meson condensation problem.
After addressing this main point,
we also discuss the relation of our argument
with the latest detailed calculation by Lee et al. [\LJMR]
that includes up to one-loop diagrams in ChPT.

We first describe the essential feature of
the original treatment of s-wave meson condensation
based on the sigma term
of an effective Lagrangian [\KAP].
As a toy model
we use the lowest-order ChPT expansion
containing s-wave $\pi$-nucleon interaction
and further truncate this Lagrangian to the minimum
number of terms in order to illustrate our points:
$$
    \cL_{1} ={1 \over 2} \l[ -\phi (\Box+\mp^2) \phi
                 +{\gSp \over \fp^2} \phi^2 {\Nb N} \r],
\eqn{\eQa}
$$
where $\phi(x)$ and $N(x)$
are the pion and the nucleon field, respectively,
and $\fp$ is the pion decay constant.
The $\gSp$ is the $\pi$-$N$ sigma term,
$$
     \gSp ={1 \over 2}(m_u+m_d)
           \Ex<N|{{\bar u} u}+{{\bar d} d}|N>.
\eqn{\eQaa}
$$
For $\cL_{1}$,
the $\pi$-$N$ scattering amplitude in tree approximation
is given by
$$
     \TpN^{(1)} ={\gSp \over \fp^2}.
\eqn{\eQb}
$$
To estimate the effective pion mass $\mp^*$
in nuclear matter,
we may use the mean-field approximation and
replace the nucleon operator $\Nb N$ in {\eQa}
with the nuclear matter density $\rho$.
Then the pion dispersion relation becomes
$\omega^2-\bk^2-\mp^2+{\rho\cdot\gSp / \fp^2} =0$.
The effective pion mass $\mp^*$ is defined
by $\mp^* \equiv \omega$($\bk=0$),
and the critical density $\rho_c$
for pion condensation is determined from the condition
$m_\pi^*=0$.
In the present case we obtain
$$
     [\mp^*(1)]^2 =\mp^2 -\rho {\gSp \over \fp^2}.
\eqn{\eQc}
$$
and
$$
     \rho_c ={\mp^2 \fp^2 \over \gSp}.
\eqn{\eQd}
$$

The second approach used in [\YNK, \YNMK]
may be summarized as follows.
One defines the pion extrapolating field $\pi(x)$ by
$$
     \pi(x) \equiv {1 \over \mp^2 \fp}
              \partial_\mu A^\mu(x)
            = {m_q \over \mp^2 \fp}
              {\bar q}(x) \gamma_5 q(x),
\eqn{\eQe}
$$
where $A^\mu(x)$ is the axial current,
and the last equality is given by QCD.
Due to its ``simplicity"
this definition of the pion field is frequently
used in QCD, the NJL model and the non- linear sigma model
$\pi^a (x)$ = ${\rm Tr}(\tau^a  U(x))$.
With this operator $\pi(x)$,
the $\pi$-$N$ scattering amplitudes
for on- and off-shell momenta of the pions are
``defined'' by
$$
     \TpN^{(2)} =i^2 (\mp^2-(k')^2) (\mp^2-k^2)
                {\int d^4x d^4y\,} e^{ik'x}e^{-iky}
                            \Ex<N'| T \pi(x) \pi(y) |N>,
\eqn{\eQf}
$$
where $k$ ($k'$) is the incoming (outgoing) pion momentum.
The amplitude $\TpN$ in {\eQf} satisfies
the Adler condition and, at the Weinberg point,
it also satisfies the well-known relation
with the sigma term
[\ADLa, \BRO].
For forward scattering,
the general form of $\TpN$
that is consistent with the low-energy theorems
can be written as
$$
     \TpN^{(2)} =
     {k^2 +(k')^2-\mp^2 \over \fp^2 \mp^2} \gSp +\TpN',
\eqn{\eQg}
$$
where only the $\gSp$-dependent terms are explicitly shown;
these terms become identical to the amplitude
in eq.{\eQb} for on-mass-shell mesons
[\COMa].
The remaining term, $\TpN'$, contains the Born terms,
the Weinberg-Tomozawa term, etc.,
and gives important contributions to the on-shell $\pi$-$N$
scattering amplitude
[\DEE, \YNK, \YNMK, \LUTa, \COMb].
However, here we neglect these terms
in order to concentrate on the role of the sigma term.
Applying the mean field approximation,
the $\mp^*$ that corresponds
to the $\pi$-$N$ amplitude eq.{\eQg} is found to be
$$
     [\mp^*(2)]^2 =\mp^2 {1 +\rho {\gSp \over \mp^2\fp^2}
                                  \over
                         1+2 \rho {\gSp \over \mp^2\fp^2}}.
\eqn{\eQi}
$$
At very low nuclear densities $\mp^*(1) \approx \mp^*(2)$
but, for larger $\rho$, $\mp^*(1)$ and $\mp^*(2)$
behave very differently.
In particular, eq.{\eQi} tends to $\mp/\sqrt{2}$ as
$\rho {\gSp \over \mp^2\fp^2}  \rightarrow \infty$,
rendering meson condensation highly unlikely.

The difference between $\mp^*$ of eq.{\eQi}
and $\mp^*$ of eq.{\eQc} represents
the gist of the current controversy
on meson condensation.
In view of the great phenomenological success
of ChPT and the PCAC approaches,
it is puzzling that their predictions on $\mp^*$,
as they stand, differ so drastically
[\COMc].
In what follows we clarify the origin of this discrepancy
and show that $\cO(\rho^2)$ terms are necessary
to resolve the problem.

Before going into the specificity,
we first recall a general argument.
For a given Lagrangian $\cL$, the finite-density
pion Green function is defined by
$$
G_{\rho}(x;\varphi) =
\Ex<\rho| T \varphi(x) \varphi(0) |\rho>,
\eqn{\eQla}
$$
where $\ket|\rho>$ is the ground state
(with baryon density $\rho$)
of the system governed by $\cL$,
and $\varphi(x)$ is an arbitrary operator
for the pion field.
The field $\varphi$ can be anything so long as
it connects one-pion state to vacuum, i.e.,
$\Ex<\pi|\varphi(x)|0> \neq 0$.
The pole position of $G_{\rho}(x;\varphi)$
corresponds to the energy $E_n$
of a pionic-mode intermediate state $\ket|n>$
that can be connected to $\ket|\rho>$ via $\varphi$.
Note that $E_n$, which is determined by $\cL$ itself,
is {\it independent}
of the choice of $\varphi$.
It then follows that $\mp^*$,
which is uniquely given
by the pole position of $G_{\rho}(x;\varphi)$,
must be independent of $\varphi$.

Now, the two amplitudes
$\TpN^{(1)}$ eq.{\eQb} and $\TpN^{(2)}$ eq.{\eQg},
although identical on the mass shell,
exhibit completely different off-mass-shell behaviors.
As is well known,
the off-mass-shell values of
the $\pi$-$N$ amplitudes depend on
the choice of the extrapolating field.
In our case the difference
between $\TpN^{(1)}$ and $\TpN^{(2)}$
reflects the two non-equivalent extrapolating fields,
$\phi(x)$ [eq.{\eQa}] and
$\pi(x)$ [eq.{\eQe}].
Naively, one might ascribe the
variance between $\mp^*(1)$ and $\mp^*(2)$
to the different off-mass-shell behaviors of the $\pi$-$N$
scattering amplitudes.
This interpretation, however, is invalidated
by the above general argument;
even if $\pi$-$N$ scattering amplitudes
exhibit different off-mass-shell behaviors
corresponding to different extrapolation fields,
$\mp^*$ itself should remain unaffected
insofar as the Lagrangian of the system is held fixed.
Therefore, it is not appropriate to attribute
the discrepancy between $\mp^*(1)$ and $\mp^*(2)$
to the off-mass-shell problem.

To gain more insight into the nature of this difference,
we consider an effective Lagrangian $\cL_2$
which, at the tree level, reproduces
the first term of the $\pi$-$N$ scattering amplitude {\eQg}
and leads to the effective mass eq.{\eQi}:
$$
    \cL_{2} ={1 \over 2} \l[ -\pi (\Box+\mp^2) \pi
                         -{\gSp \over \fp^2}
           (\pi^2 +{2 \over \mp^2}\pi \Box \pi) {\Nb N} \r].
\eqn{\eQh}
$$
$\cL_2$ differs from $\cL_1$ [eq.{\eQa}]
by the existence of the interaction term that involves
$\Box\pi$ (``box term'')
[\COMd].
Since the meson field in ChPT
is nothing more than an integration variable and
has no physical meaning by itself
[\LEU, \GSS],
it is useful to examine here
to what extent $\cL_2$ can be transformed
into $\cL_1$ via a meson field redefinition.
To this end, we apply the mean field approximation,
${\Nb N} \rightarrow \rho$, to {\eQh}
[\COMe]
and introduce a new meson field $\tilde{\phi}(x)$
defined by
$$
   \pi(x)=
    \l( 1 -\rho {\gSp \over \fp^2\mp^2} \r)
    \tilde{\phi}(x).
\eqn{\eQj}
$$
With $\tilde{\phi}(x)$, the Lagrangian {\eQh}
can be rewritten as
$$
    \cL_2 =\l( 1 -\rho {\gSp \over \fp^2\mp^2} \r)^2
       {1 \over 2} \l[ -\tilde{\phi}
                      (\Box+\mp^2) \tilde{\phi}
          -{\gSp \over \fp^2}
          (\tilde{\phi}^2 + {2 \over \mp^2}
         \tilde{\phi} \Box \tilde{\phi}) \rho \r].
\eqn{\eQk}
$$
Expanding this in $\rho$, we obtain
$$
    \cL_2 ={1 \over 2}
    \l[ -\tilde{\phi} (\Box+\mp^2) \tilde{\phi}
     + \rho {\gSp \over \fp^2} \tilde{\phi}^2 \r]
     +\cO(\rho^2).
\eqn{\eQl}
$$
This Lagrangian is identical to eq.{\eQa}
($\phi \leftrightarrow \tilde{\phi}$),
if the terms of $\cO(\rho^2)$ are neglected,
a feature which is in accord with the fact
that $\mp^*(1) =\mp^*(2)$
up to order $\rho$.

The equivalence of
$\cL_1$ and $\cL_2$ breaks down at the $\cO(\rho^2)$ level,
and this non-equivalence
is responsible for the difference between
$\mp^*(1)$ and $\mp^*(2)$.
Although this statement itself is correct,
the real significance of this statement
hinges upon the question:
Can the existing formalisms
make a meaningful distinction
between $\mp^*(1)$ and $\mp^*(2)$ ?
For the sake of clarity, we rephrase this question
by referring back to the general discussion of
$G_{\rho}(x;\varphi)$ [eq.{\eQla}].
For the Lagrangian $\cL_2$,
one can consider two Green functions,
$G^{(2)}(x;\pi)  \equiv  G_{\rho}(x;\varphi=\pi)$
and
$ G^{(2)}(x;\tilde{\phi}) \equiv
G_{\rho}(x;\varphi=
[1 -(\rho \gSp /\fp^2\mp^2)] \tilde{\phi})$.
These Green functions are not identical
but their pole positions give the same effective mass,
$\mp^*(2)$ of eq.{\eQi}.
On the other hand, if we consider the Green function
$G^{(1)}(x;\phi) \equiv G_{\rho}(x;\varphi=\phi)$
governed by $\cL^{(1)}$,
with $\phi$ being the field appearing in {\eQa},
the pole position will move to
$\mp^*(1)$ [eq.{\eQc}],
reflecting the change of the basic Lagrangian.
Now, if one has a definite criterion to decide which
effective Lagrangian, $\cL^{(1)}$ or $\cL^{(2)}$,
is superior,
then one would know which effective mass to use,
$\mp^*(1)$ or $\mp^*(2)$.
So, the crucial question is whether
the formalisms so far developed allow us to
decide which of $\cL^{(1)}$ and $\cL^{(2)}$
is a better choice.

{}From the ChPT point of view,
one might assert that,
to a given chiral order,
$\cL^{(1)}$ is unique (modulo field transformations)
and hence any other Lagrangians,
including $\cL^{(2)}$,
that deviate therefrom within the same chiral order
should be discarded.
The issue, however, is more subtle.
We have shown in {\eQl} that
$\cL^{(1)}$ and $\cL^{(2)}$ differ by
terms of $\cO(\rho^2)$.
However, since $\cL^{(1)}$ is devoid of
terms containing $(\Nb N)^2$ like $\pi^2 (\Nb N)^2$
(which in the mean-field approximation would
give contributions of $\cO(\rho^2)$),
it goes beyond the accuracy of $\cL^{(1)}$
to discuss the difference of $\cO(\rho^2)$
between $\cL^{(1)}$ and $\cL^{(2)}$.
If $\cL^{(1)}$ were a fundamental Lagrangian,
one might still be able to justify the absence of terms
involving $(\Nb N)^n$ ($n \geq 2$) in {\eQa}
[\COMf].
However, $\cL^{(1)}$ being an effective Lagrangian,
one cannot a priori exclude from $\cL^{(1)}$
terms containing $(\Nb N)^n$ ($n \geq 2$)
[\COMg].
Therefore, there is no compelling reason
to prefer $\cL^{(1)}$ to $\cL^{(2)}$.

Meanwhile, from the current-algebra-plus-PCAC viewpoint,
one might claim that $\cL^{(2)}$ is a ``natural" choice,
and that $\cL^{(1)}$ is an approximate Lagrangian
obtained from $\cL^{(2)}$
by ignoring the $\cO(\rho^2)$ terms in eq. {\eQl}.
However, this argument is subject
to the same criticism as above;
therefore $\cL^{(2)}$ cannot be considered as
a better approximation than $\cL^{(1)}$.

These observations make it clear
that the true understanding of
the difference between $\mp^*(1)$ and $\mp^*(2)$
requires a knowledge of terms of $\cO(\rho^2)$
in the effective Lagrangian itself.
In other words,
the discrepancy between $\mp^*(1)$ and $\mp^*(2)$
represents the effects of
two (or more) -nucleon interaction terms
which have not been addressed so far.
This is a new type of matter effect.
Usually, matter effects of $\cO(\rho^2)$
such as the Lorentz-Lorenz-Ericson-Ericson effect,
the in-medium modifications of $g_A$, $m_N$ etc.,
are regarded as well-defined corrections
to the linear-density approximation for ChPT.
However, the above discussion shows that
there exists a class of matter effects
which arise from higher-order density terms
in the effective Lagrangian,
and whose form
vary according to the extrapolating field.
We re-emphasize that this extrapolating-field dependence
does not affect $\mp^*$
if no truncation is introduced to the chiral effective
Lagrangian, and if $G_{\rho}(x;\varphi)$ is treated exactly.
It is only when an approximation is introduced either
in $\cL$ or in the calculation of $G_{\rho}(x;\varphi)$
that the resulting $\mp^*$ becomes dependent on
the interpolating field.

So far we concentrated on the sigma term
in the pion sector.
The situation is essentially the same for the kaon case.
We now discuss the meaning of our argument
in the light of the recent developments
in the ChPT approach to kaon condensation
[\BLRT,\LJMR].
The basic problems with the earlier ChPT calculations
[\KAP-\BKRT] were:
(i) chiral-counting was not done consistently;
(ii) the $K$-$N$ scattering amplitudes
did not possess a correct energy dependence
to reproduce the scattering data.

Regarding problem (i),
a systematic ChPT calculation that respects
chiral-order counting
based on the heavy-fermion formalism
has been carried out to the tree order
by Brown, Lee, Rho and Thorsson [\BLRT],
and to the one-loop order by Lee, Jung, Min and Rho
[\LJMR, \BKM].
As far as the ordinary chiral counting in vacuum
is concerned, these calculations are complete up
to the stated chiral orders,
but multiple-fermion terms do not appear
in these calculations
[\COMh].
This gives the impression
that we need not worry about the absence
of multi-fermion terms in $\cL^{(1)}$.
However, because a finite-density system has
an additional scale $\rho$,
chiral counting in nuclear matter is not
as firmly established as in vacuum.
This caveat becomes particularly important
in applying ChPT to systems of higher densities.
Thus, a further study is needed
to check whether the contributions of
multiple-fermion terms are as suppressed as
the ordinary chiral counting would indicate
[\COMi].
Until this point is settled,
it seems unsafe to invoke the ordinary chiral counting
to justify ignoring $\cO(\rho^2)$ terms
that are responsible for
the difference between $\cL^{(1)}$ and $\cL^{(2)}$
[\COMj].
Even if one insists on the ordinary chiral counting,
the inclusion of meson loops in the ChPT
calculations requires, to the same chiral order,
that at least two nucleon terms, $\cO(\rho^2)$ terms,
should in general be taken into account in calculating
the Green function in nuclear matter.
In this sense also,
$\cO(\rho^2)$ terms like $\pi^2 (\Nb N)^2$
need to be included in the Lagrangian itself.

As for problem (ii),
Lee et al.[\LJMR] considered the energy dependence
coming from the one loop diagrams and the
resonance $\Lambda^*(1405)$,
and were able to reproduce reasonably well
the existing data
on the s-wave $K$-$N$ scattering amplitude.
The pronounced energy dependence
in the s-wave $\bar{K}$-$N$ ($I=0$) scattering amplitude
was reproduced by adjusting the parameters characterizing
$\Lambda^*$.
However, the limited precision of
the present experimental data
hinders an accurate test of
the energy dependence due to the loop diagrams;
this dependence should be most visible
in the $K$-$N$ ($I=1$) channel.
Now, since kaon condensation
depends on the energy behavior of the $K$-$N$
amplitudes from threshold ($\omega= m_K$)
down towards $\omega=0$,
it is important to know whether
this subthreshold energy behavior is reproduced accurately
by the one-loop corrections.
In the language of the empirical low-energy expansion,
$$
    T_{KN} = a + b (\omega^2 - m_K^2)
 + \cO ( (\omega^2 - m_K^2)^2) ,
\eqn{\eQn}
$$
this means that the parameters in $\cL$ must
account not only for the s-wave $K$-$N$ scattering length
$a$
but also for the s-wave effective range $b$ [\COMk].
This is at present a difficult task
due to the paucity of experimental data.
In the phenomenological approach of [\YNMK]
this difficulty is reflected in the fact
that the $\Sigma_{KN}$ was treated
as a parameter.
Further experimental information
on low-energy $K$-$N$ scattering as well as
calculations that include $\cO(\rho^2)$ terms
are required to make progress in this discussion.

Finally, there is a possibility of using
astrophysical input to place constraints
on the role of the multi-fermion terms
in the effective chiral Lagrangian.
According to Brown and Bethe
[\BB],
the kaon condensate, if it exists, should
lead to the formation of ``nuclear star" matter
(instead of the neutron star matter) and
the proliferation of pygmy blackholes.
If observational support for this scenario becomes
compelling enough,
that may be construed as indirect evidence
for the essential correctness of
the effective Lagrangian so far used in
the ChPT approach to kaon condensation.

\vskip 0.8cm

The authors wish to express their sincere thanks
to Mannque Rho for his many illuminating remarks and
also for communicating the results of ref.[\LJMR] prior
to publication.
KK gratefully acknowledges the useful discussions with
A. Wirzba, V. Thorsson, U.-G. Meissner,
N. Kaiser, C.-H. Lee and T.-S. Park
at the Workshop on Chiral Symmetry
 in Hadron and Nuclei
at the ECT$^*$, Trento.

\vskip 0.5mm

\pagebreak
\par \penalty-400 \vskip\chapterskip
   \spacecheck\referenceminspace \immediate\closeout\referencewrite
   \line{\hfil REFERENCES\hfil}\vskip\headskip
   \catcode`@=11
   \input reference.aux
   \catcode`@=12
\bye